\documentclass[12pt]{article}
\usepackage{amssym}

\def\R{\mbox{$\Bbb R$}}
\def\N{\mbox{$\Bbb N$}}
\def\arcsinh{\mathop{\rm arcsinh}\nolimits}
\def\case#1#2{{\textstyle{#1\over #2}}}
\def\Ab{\bar{A}}
\def\cosech{\mathop{\rm cosech}\nolimits}

\title{
PT-SYMMETRIC SQUARE WELL AND THE ASSOCIATED SUSY HIERARCHIES}
\author{B. BAGCHI\thanks{E-mail: bbagchi@cucc.ernet.in} \ and S.
MALLIK \\
{\small \sl Department of Applied Mathematics, University of Calcutta,} \\
{\small \sl 92 Acharya Prafulla Chandra Road, Calcutta 700 009, India}\\[10pt] 
C. QUESNE\thanks{Directeur de recherches FNRS; E-mail: cquesne@ulb.ac.be} \\
{\small \sl Physique Nucl\'eaire Th\'eorique et Physique Math\'ematique,}\\ {\small \sl
Universit\'e Libre de Bruxelles, Campus de la Plaine CP229,} \\ {\small \sl  Boulevard~du
Triomphe, B-1050 Brussels, Belgium}}
\date{ }
\begin{document}
\baselineskip=22pt plus 1pt minus 1pt
\maketitle

\begin{abstract} 
The PT-symmetric square well problem is considered in a SUSY framework. When the
coupling strength $Z$ lies below the critical value $Z_0^{\rm (crit)}$ where PT symmetry
becomes spontaneously broken, we find a hierarchy of SUSY partner potentials, depicting
an unbroken SUSY situation and reducing to the family of $\sec^2$-like potentials in the
$Z \to 0$ limit.  For $Z$ above $Z_0^{\rm (crit)}$, there is a rich diversity of SUSY
hierarchies, including some with PT-symmetry breaking and some with partial
PT-symmetry restoration.
\end{abstract}

\vspace{1cm}
\noindent
Running head: PT-symmetric square well

\noindent
PACS: 03.65.Fd, 03.65.Ge

\noindent
Keywords: supersymmetric quantum mechanics, PT symmetry, square well
\newpage
%
%
\section{Introduction}

One of the simplest solvable one-dimensional potentials in quantum mechanics is the
square well potential~\cite{schiff}. It describes the confinement of a particle trapped in a
box with infinite walls. As is well known, the energy spectrum of such a particle is entirely
discrete and nondegenerate.\par
%
%
The square well potential problem has been examined in a supersymmetric (SUSY) context
too~\cite{sukumar}. One finds that a sequence of Hamiltonians can be generated starting
from a free-particle potential inside the well. Such a hierarchy, which is controlled by a
family of $\sec^2$-like potentials, enjoys the property that its adjacent members are
SUSY partners.\par
%
%
Recently, the idea of PT symmetry in quantum mechanics has evoked a lot of
interest~\cite{bender, fernandez, cannata, delabaere, bagchi00a, bagchi00b}, especially
in connection with its nontrivial role in non-Hermitian SUSY systems~\cite{andrianov,
bagchi00c, znojil00, bagchi01a}. Briefly, PT-symmetric Hamiltonians are the ones which
are invariant under joint action of parity (P: $x \to -x$) and time reversal (T: ${\rm i} \to
- {\rm i}$). More importantly, such Hamiltonians are conjectured~\cite{bessis, kretschmer,
bagchi01b, japaridze, mosta, ahmed} to preserve the reality of their bound-state
eigenvalues, except possibly when PT is spontaneously broken. SUSY methods enable one
to construct non-Hermitian Hamiltonians with real and/or complex discrete eigenvalues by
complexifying the underlying superpotential. Indeed this has allowed extensions of a
number  of exactly solvable models in ordinary quantum mechanics to the non-Hermitian
sector~\cite{andrianov, bagchi01a}.\par
%
%
In this Letter, our primary concern is to derive a PT-analogue of the aforementioned
$\sec^2$-hierarchy. In this regard, the PT-symmetric version of the square well
potential and its associated SUSY partner prove to be our natural starting point.
Enquiries into the functioning of the PT-symmetric square well reveal~\cite{znojil01a,
znojil01b} that a critical value of the coupling parameter exists beyond which some energy
levels appear in complex-conjugate pairs. This signals an onset of a PT-spontaneously
broken phase. However, below the critical value, PT symmetry is unbroken with all the
energy levels remaining real. As we plan to show, the existence of a PT-symmetry
broken phase is responsible for some new features in the SUSY extension to
non-Hermitian Hamiltonians.\par
%
%
\section{PT-Symmetric Square Well}

A PT-symmetric square well potential on the interval $(-1, 1)$ can be defined in the
manner~\cite{znojil01a}
\begin{equation}
  V_{\rm R}(x) = - {\rm i} Z, \qquad V_{\rm L}(x) = {\rm i} Z,  \label{eq:SW}
\end{equation}
where $Z$ is the coupling strength and L (R) denotes the region $-1 < x < 0$ ($0 < x <
1$). The wave functions, at the end points of the interval, are enforced to be vanishing as
is normally the case with the real square well problem with impenetrable walls.\par
%
%
{}For real eigenvalues $E_n$ ($n=0$, 1, 2,~\ldots), the Schr\"odinger equation for
(\ref{eq:SW}) is equivalent to the pair
\begin{equation}
  \psi''_{n{\rm R}} = \kappa_n^2 \psi_{n{\rm R}}, \qquad \psi''_{n{\rm L}} =
  \kappa_n^{*2} \psi_{n{\rm L}},  \label{eq:SE}
\end{equation}
where
\begin{equation}
 \kappa_n^2 = - E_n - {\rm i}Z = (s_n - {\rm i}t_n)^2, \qquad s_n, t_n \in \R.
  \label{eq:kappa}  
\end{equation}
Solving (\ref{eq:kappa}) gives
\begin{equation}
  t_n = \frac{1}{\sqrt{2}} \left(E_n + \sqrt{E_n^2 + Z^2}\right)^{1/2}, \qquad
  s_n = \frac{Z}{\sqrt{2}} \left(E_n + \sqrt{E_n^2 + Z^2}\right)^{-1/2}.
\end{equation}
\par
%
%
The solutions of (\ref{eq:SE}) fulfilling the conditions $\psi_{n{\rm R}}(1) = \psi_{n{\rm
L}}(-1) = 0$ may be written as
\begin{equation}
  \psi_{n{\rm R}} = C_{n{\rm R}} \sinh[\kappa_n(1-x)], \qquad \psi_{n{\rm L}} =
  C_{n{\rm L}} \sinh[\kappa_n^*(1+x)].  \label{eq:SE-sol}  
\end{equation}
\par
%
%
To establish a link between the (complex) constants $C_{n{\rm R}}$ and $C_{n{\rm
L}}$, we see that the continuity of the wave function and its derivative at $x=0$ imposes
the conditions
\begin{equation}
  \kappa_n \coth \kappa_n + \kappa_n^* \coth \kappa_n^* = 0,  \label{eq:cond-0-1} 
\end{equation}
\begin{equation}
  \frac{C_{n{\rm R}}}{C_{n{\rm L}}} = \frac{\sinh \kappa_n^*}{\sinh \kappa_n}.
  \label{eq:cond-0-2}
\end{equation}
Now, if we require $\psi_{n{\rm R}}$ and $\psi_{n{\rm L}}$ to be also PT-symmetric
near the origin, that is
\begin{equation}
  \psi_{n{\rm R}}(0) = \psi_{n{\rm L}}(0) = \alpha_n, \qquad \partial_x \psi_{n{\rm
  R}}(0) = \partial_x \psi_{n{\rm L}}(0) = {\rm i} \beta_n,  \label{eq:PT-0} 
\end{equation}
where the parameters $\alpha_n$, $\beta_n \in \R$, then use of (\ref{eq:PT-0}) enables
us to rewrite the eigenfunctions (\ref{eq:SE-sol}) as 
\begin{equation}
  \psi_{n{\rm R}} = \frac{\alpha_n}{\sinh \kappa_n} \sinh[\kappa_n (1-x)], \qquad 
  \psi_{n{\rm L}} = \frac{\alpha_n}{\sinh \kappa_n^*} \sinh[\kappa_n^* (1+x)].
  \label{eq:wf}
\end{equation}
\par
%
%
It is to be noted that the condition (\ref{eq:cond-0-1}), which is independent  of
$C_{n{\rm R}}$ and $C_{n{\rm L}}$, can also be put in the form
\begin{equation}
  s_n \sinh 2s_n + t_n \sin 2t_n = 0,  \label{eq:cond-0-1-bis} 
\end{equation}
where we have used (\ref{eq:kappa}). Setting $s_n \sinh 2s_n = 2 \sinh^2 S_n$ and
$t_n = \frac{\pi}{2} T_n$, we can recast (\ref{eq:cond-0-1-bis}) as 
\begin{equation}
  4 \sinh^2 S_n = - \pi T_n \sin \pi T_n.  \label{eq:cond-0-1-ter}
\end{equation}
Equation (\ref{eq:cond-0-1-ter}) shows that the points $(T_n, S_n)$ of the $T$-$S$ plane
corresponding to the eigenvalues $E_n$ belong to the $Z$-independent curve
\begin{equation}
  S = X(T) = \arcsinh \case{1}{2} \sqrt{- \pi T \sin\pi T},  \label{eq:X}  
\end{equation}
where $2m-1 < T < 2m$, $m=1$, 2,~\ldots. On the other hand, Eq.~(\ref{eq:kappa})
may be exploited to obtain $\sinh^2 S_n = \frac{Z}{2\pi T_n} \sinh \frac{2Z}{\pi T_n}$,
which in turn exposes the $Z$-dependence of $S$:
\begin{equation}
  S = Y(Z, T) = \arcsinh \sqrt{\frac{Z}{2\pi T} \sinh \frac{2Z}{\pi T}}.  \label{eq:Y}
\end{equation}
\par
%
%
The pair of equations (\ref{eq:X}) and (\ref{eq:Y}) imply that the points $(T_n, S_n)$,
and in consequence the energies $E_n$, are at the intersections of the two curves $S =
X(T)$ and $S = Y(Z, T)$. As shown in~\cite{znojil01a, znojil01b}, if $Z$ is below the
critical threshold $Z_0^{\rm (crit)} \approx 4.48$, there are two real eigenvalues in every
interval $2\nu-1 < T < 2\nu$, where $\nu=1$, 2,~\ldots. At $Z = Z_0^{\rm (crit)}$, the
two lowest real eigenvalues $E_0$ and $E_1$ merge and for $Z > Z_0^{\rm (crit)}$ move
into the complex plane, where they become complex conjugate. The corresponding
eigenfunctions $\psi_0$ and $\psi_1$ then cease to be PT-symmetric. More generally,
there exists a naturally-ordered increasing sequence of critical couplings $Z_0^{\rm
(crit)} < Z_1^{\rm (crit)} < \cdots < Z_{\nu}^{\rm (crit)} < \cdots$ at which a pair of
real eigenvalues $(E_{2\nu}, E_{2\nu+1})$ merges and moves into the complex plane.
\par
%
%
In Secs.~2--5, we shall consider the unbroken PT-symmetry case ($Z < Z_0^{\rm
(crit)}$), leaving the discussion of the broken PT-symmetry one ($Z > Z_0^{\rm
(crit)}$) for Sec.~6.\par
%
%
\section{SUSY Partner in the Unbroken PT-Symmetry Case}

With the above preliminaries on the PT-symmetric square well potential, we proceed to
the construction of its SUSY partner in the $Z < Z_0^{(\rm crit)}$ case. Adopting the
notations of~\cite{bagchi01a}, we consider the following factorization scheme
\begin{equation}
  H^{(\pm)} = - \frac{d^2}{dx^2} + V^{(\pm)}(x) - E_0 \equiv (\Ab A, A \Ab),
  \label{eq:Hpm}
\end{equation}
corresponding to an arbitrary factorization energy $E = E_0$. In (\ref{eq:Hpm}),
$V^{(\pm)}$ are the SUSY partner potentials and the operators $A$, $\Ab$ may be
defined in terms of a superpotential $W(x)$ as
\begin{equation}
  A = \frac{d}{dx} + W(x), \qquad \Ab = - \frac{d}{dx} + W(x).  \label{eq:A}
\end{equation}
Inserting (\ref{eq:A}) in (\ref{eq:Hpm}) gives
\begin{equation}
  V^{(\pm)} = W^2 \mp W' + E_0.
\end{equation}
\par
%
%
Identifying $V^{(+)}$ with the square well potential (\ref{eq:SW}), that is, taking
$V^{(+)}_{\rm R} = - {\rm i}Z$ and $V^{(+)}_{\rm L} = {\rm i}Z$, we obtain for
$W(x)$ the following differential equations
\begin{equation}
  W_{\rm R}^2 - W'_{\rm R} = \kappa_0^2, \qquad W_{\rm L}^2 - W'_{\rm L} =
  \kappa_0^{*2}.  \label{eq:W-diff} 
\end{equation}
Solving (\ref{eq:W-diff}) we get
\begin{equation}
  W_{\rm R}(x) = - \kappa_0 \tanh[\kappa_0 (x-x_{\rm R})], \qquad  W_{\rm L}(x) = -
  \kappa_0^* \tanh[\kappa_0^* (x+x_{\rm L})],  \label{eq:W-sol}
\end{equation}
where $x_{\rm R}$ and $x_{\rm L}$ are two integration constants.\par
%
%
{}From (\ref{eq:W-sol}) it results that the partner potential to (\ref{eq:SW}) is given by
\begin{equation}
  V^{(-)}_{\rm R}(x) = - {\rm i}Z + 2 \kappa_0^2 \cosech^2[\kappa_0 (1-x)], \quad
  V^{(-)}_{\rm L}(x) = {\rm i}Z + 2 \kappa_0^{*2} \cosech^2[\kappa_0^* (1+x)].
  \label{eq:V-}
\end{equation}
The partner Hamiltonian $H^{(-)}$ is therefore PT-symmetric. Note that in (\ref{eq:V-}),
$x_{\rm R}$ and $x_{\rm L}$ have been chosen as $x_{\rm R} = 1 - {\rm i} \pi/(2
\kappa_0)$ and $x_{\rm L} = 1 - {\rm i} \pi/(2\kappa_0^*)$ to ensure that
$V^{(-)}(x)$ blows up at the end points $x=-1$ and $x=+1$, which is as it should be.
Observe that the superpotential $W(x)$ also blows up at these points:
\begin{equation}
  W_{\rm R}(x) = \kappa_0 \coth[\kappa_0 (1-x)], \qquad
  W_{\rm L}(x) = - \kappa_0^* \coth[\kappa_0^* (1+x)].
  \label{eq:W}
\end{equation}  
\par
%
%
To obtain the eigenfunctions associated with $H^{(-)}$ we first remark that the ground
state eigenfunction $\psi^{(+)}_0(x)$ of $H^{(+)}$, given by (\ref{eq:wf}) when $n=0$,
is annihilated by the operator $A$:
\begin{equation}
  \left[\frac{d}{dx} + W_{\rm R}(x)\right] \psi^{(+)}_{0{\rm R}}(x) =
  \frac{\alpha^{(+)}_0}{\sinh \kappa_0} \left\{\frac{d}{dx} + \kappa_0 \coth[\kappa_0
  (1-x)]\right\} \sinh[\kappa_0 (1-x)] = 0,  \label{eq:A-wf0}
\end{equation}
where the superscript $(+)$ is appended to the eigenfunction and the coefficient
parameter to signify that we are dealing with the $H^{(+)}$ component. A similar result
like (\ref{eq:A-wf0}) holds for $\left[\frac{d}{dx} + W_{\rm L}(x)\right]
\psi^{(+)}_{0{\rm L}}(x)$.\par
%
%
Exploiting then the intertwining character of SUSY, the eigenfunctions $\psi^{(-)}_n(x)$,
$n=0$, 1, 2,~\ldots, of $H^{(-)}$ are obtained by application of $A$ on
$\psi^{(+)}_{n+1}$ subject to the preservation of the boundary and continuity
conditions:
\begin{eqnarray}
  \psi^{(-)}_{n{\rm R}}(1) & = & 0,  \qquad \psi^{(-)}_{n{\rm L}}(-1) = 0,
         \label{eq:partner-bound} \\
  \psi^{(-)}_{n{\rm R}}(0) & = & \psi^{(-)}_{n{\rm L}}(0), \qquad \partial_x
         \psi^{(-)}_{n{\rm R}}(0) = \partial_x \psi^{(-)}_{n{\rm L}}(0). 
         \label{eq:partner-cont}
\end{eqnarray}
We get in this way
\begin{eqnarray}
  \psi^{(-)}_{n{\rm R}}(x) & = & C^{(-)}_{n{\rm R}} \frac{\alpha^{(+)}_{n+1}}{\sinh
        \kappa_{n+1}} \sinh[\kappa_{n+1} (1-x)] \nonumber \\
  && \mbox{} \times \{- \kappa_{n+1} \coth[\kappa_{n+1} (1-x)] + \kappa_0
        \coth[\kappa_0 (1-x)]\}, \nonumber \\
  \psi^{(-)}_{n{\rm L}}(x) & = & C^{(-)}_{n{\rm L}} \frac{\alpha^{(+)}_{n+1}}{\sinh
        \kappa_{n+1}^*} \sinh[\kappa_{n+1}^* (1+x)] \nonumber \\
  && \mbox{} \times \{\kappa_{n+1}^* \coth[\kappa_{n+1}^* (1+x)] - \kappa_0^*
        \coth[\kappa_0^* (1+x)]\}, 
\end{eqnarray}
where $C^{(-)}_{n{\rm R}}$ and $C^{(-)}_{n{\rm L}}$ are (complex) constants. These
eigenfunctions satisfy the boundary conditions (\ref{eq:partner-bound}) because of the
limiting relations of the type $\lim_{x\to 1} \{\sinh[\kappa_{n+1} (1-x)]/ \sinh[\kappa_0
(1-x)]\} = \kappa_{n+1}/\kappa_0$.\par
%
%
On the other hand, because of the continuity conditions (\ref{eq:partner-cont}) it turns
out that
\begin{equation}
  C^{(-)}_{n{\rm R}} = C^{(-)}_{n{\rm L}} = C^{(-)}_n,  \label{eq:C-}
\end{equation}
along with
\begin{equation}
  \kappa_{n+1}^2 - \kappa_{n+1}^{*2} = \kappa_0^2 - \kappa_0^{*2} = - 2 {\rm i}Z, 
\end{equation}
where we have used (\ref{eq:kappa}).\par
%
%
As previously, if we require $\psi^{(-)}_n$ to be also PT-symmetric near the origin and so
denote by $\alpha^{(-)}_n$ and ${\rm i} \beta^{(-)}_n$ the respective values of
$\psi^{(-)}_n(x)$ and $\partial_x \psi^{(-)}_n(x)$ at $x=0$, where $\alpha^{(-)}_n,
\beta^{(-)}_n \in \R$, we obtain, on account of (\ref{eq:C-}), the following forms of the
eigenfunctions of $H^{(-)}$:
\begin{eqnarray}
  \psi^{(-)}_{n{\rm R}}(x) & = & \frac{\alpha^{(-)}_n \sinh[\kappa_{n+1} (1-x)]}
        {\sinh\kappa_{n+1} (\kappa_{n+1} \coth \kappa_{n+1} - \kappa_0 \coth
        \kappa_0)} \nonumber \\
  && \mbox{} \times \{\kappa_{n+1} \coth[\kappa_{n+1} (1-x)] - \kappa_0
        \coth[\kappa_0 (1-x)]\}, \nonumber \\
  \psi^{(-)}_{n{\rm L}}(x) & = & \frac{\alpha^{(-)}_n \sinh[\kappa_{n+1}^* (1+x)]}
        {\sinh\kappa_{n+1}^* (\kappa_{n+1}^* \coth \kappa_{n+1}^* - \kappa_0^* \coth
        \kappa_0^*)} \nonumber \\
  && \mbox{} \times \{\kappa_{n+1}^* \coth[\kappa_{n+1}^* (1+x)] - \kappa_0^*
        \coth[\kappa_0^* (1+x)]\},  \label{eq:partner-wf}
\end{eqnarray}
where $\alpha^{(-)}_n = C^{(-)}_n \alpha^{(+)}_{n+1} (- \kappa_{n+1} \coth
\kappa_{n+1} + \kappa_0 \coth \kappa_0)$.\par
%
%
We remark that the above eigenfunctions of $H^{(-)}$, defined by (\ref{eq:Hpm}) and
(\ref{eq:V-}), have SUSY related eigenvalues
\begin{equation}
  E^{(-)}_n = E^{(+)}_{n+1} = E_{n+1} - E_0 = \kappa_0^2 - \kappa_{n+1}^2.
  \label{eq:E-}
\end{equation}
Notice that in (\ref{eq:E-}), the coupling strength $Z$ only appears implicitly in the
$\kappa$'s. Equation (\ref{eq:E-}) reflects a typical unbroken SUSY feature: pairing of
the eigenvalues of the partner Hamiltonians with the ground state nondegenerate for
$n=0$, as shown by (\ref{eq:A-wf0}) for $\psi^{(+)}_{0{\rm R}}(x)$ and a similar
equation for $\psi^{(+)}_{0{\rm L}}(x)$.\par
%
%
\section{\boldmath $Z \to 0$ Limit}

At this stage, it is instructive to look into the $Z \to 0$ limit. In this limit,
Eq.~(\ref{eq:kappa}) becomes $\kappa_n^2 = - E_n = - t_n^2$. From the continuity
condition (\ref{eq:cond-0-1-bis}), which is now $t_n \sin 2t_n = 0$, we are led to
\begin{equation}
  t_n = (n+1) \frac{\pi}{2}, \qquad E_n = (n+1)^2 \frac{\pi^2}{4}.
\end{equation}
For the odd values of $n$ for which $\sin t_n = 0$, the other continuity condition
(\ref{eq:cond-0-2}) becomes useless because its right-hand side is indeterminate. Going
back to the square well eigenfunctions (\ref{eq:SE-sol}), which are now
\begin{equation}
  \psi^{(+)}_{n{\rm R}} = - {\rm i} C^{(+)}_{n{\rm R}} \sin[t_n (1-x)], \qquad
  \psi^{(+)}_{n{\rm L}} = {\rm i} C^{(+)}_{n{\rm L}} \sin[t_n (1+x)],
\end{equation}
it is however straightforward to see that the continuity conditions yield two solutions
\begin{eqnarray}
  C^{(+)}_{n{\rm R}} & = & - C^{(+)}_{n{\rm L}} = C^{(+)}_n, \qquad \cos t_n = 0,
          \nonumber \\
  C^{(+)}_{n{\rm R}} & = & C^{(+)}_{n{\rm L}} = C^{(+)}_n, \qquad \sin t_n = 0,
\end{eqnarray}
according to whether $n$ is even or odd. As a consequence, the eigenfunctions can be
written as
\begin{eqnarray}
  \psi^{(+)}_{2\nu}  & = & - {\rm i} C^{(+)}_{2\nu} (-1)^{\nu} \cos\left[(2\nu+1)
         \frac{\pi}{2} x\right], \nonumber \\
  \psi^{(+)}_{2\nu+1}  & = & {\rm i} C^{(+)}_{2\nu+1} (-1)^{\nu+1} \sin[(\nu+1) \pi
         x],  \label{eq:wf-real}
\end{eqnarray}
where we do not have to distinguish between the intervals $(-1, 0)$ and $(0, 1)$
anymore. The forms (\ref{eq:wf-real}) are in conformity with the known results for the
real square well~\cite{schiff}.\par
%
%
Let us now consider the SUSY partner as $Z \to 0$. Since $\kappa_0 = - {\rm i}t_0 = -
{\rm i} \frac{\pi}{2}$, the superpotentials in (\ref{eq:W}) become $W_{\rm R, L}(x) =
\frac{\pi}{2} \tan\left(\frac{\pi}{2} x\right)$. As a result, the partner potentials in
(\ref{eq:V-}), too, acquire the common form $V^{(-)}_{\rm R, L} = \frac{\pi^2}{2}
\sec^2 \left(\frac{\pi}{2}x\right)$ that coincides with the SUSY partner of the real
square well first obtained in~\cite{sukumar}.\par
%
%
{}For the partner eigenfunctions we obtain, in the $Z \to 0$ limit,
\begin{eqnarray}
  \psi^{(-)}_{2\nu}(x) & = & C^{(-)}_{2\nu} C^{(+)}_{2\nu+1}\, {\rm i} \frac{\pi}{2}
         (-1)^{\nu+1} \biggl\{(2\nu+2) \cos[(\nu+1) \pi x] \nonumber \\
  && \mbox{} + \tan\left(\frac{\pi}{2} x\right) \sin[(\nu+1) \pi x]\biggr\}, \nonumber \\
  \psi^{(-)}_{2\nu+1}(x) & = & C^{(-)}_{2\nu+1} C^{(+)}_{2\nu+2}\, {\rm i}
         \frac{\pi}{2} (-1)^{\nu+1} \biggl\{(2\nu+3) \sin\left[(2\nu+3) \frac{\pi}{2}
         x\right] \nonumber \\
  && \mbox{} - \tan\left(\frac{\pi}{2} x\right) \cos\left[(2\nu+3) \frac{\pi}{2} x
         \right]\biggr\}. \label{eq:partner-wf-real} 
\end{eqnarray}
\par
%
%
{}From the results of~\cite{sukumar}, we know that the real square well $V_1(x) =
V^{(+)}_{\rm R, L}(x)$ generates a whole family of $\sec^2$-like potentials with
increasing strengths,
\begin{equation}
  V_m(x) = V_1(x) + \frac{\pi^2}{4} m(m-1) \sec^2\left(\frac{\pi}{2} x\right), \qquad
  m = 1, 2, 3, \ldots, \label{eq:V-m}
\end{equation}
corresponding to a hierarchy of Hamiltonians, whose adjacent members are SUSY
partners. Having found  the PT-symmetric analogue of the second member of the family,
$V_2(x) = V^{(-)}_{\rm R, L}(x)$, we may now try to build counterparts of the other
members, $V_3$, $V_4$,~\ldots. Such a construction is outlined in the next section.\par
%
%
\section{SUSY Hierarchy in the Unbroken PT-Symmetry Case}

Let us define a hierarchy of partner Hamiltonians $H_m$, $m=1$, 2,~\ldots, whose first
member $H_1$ coincides with that of the PT-symmetric square well. According to this
description, we have the following set of SUSY partners
\begin{eqnarray}
  H^{(+)}_m & = & - \frac{d^2}{dx^2} + V^{(+)}_m(x) - E_{m,0} = H_m - E_{m,0} =
          \Ab_m A_m, \nonumber \\
  H^{(-)}_m & = & - \frac{d^2}{dx^2} + V^{(-)}_m(x) - E_{m,0} = H_{m+1} - E_{m,0} =
          A_m \Ab_m,
\end{eqnarray} 
where $V^{(+)}_m(x) = V_m(x)$, $V^{(-)}_m(x) = V_{m+1}(x)$, $m=1$, 2,~\ldots. For
$m=1$, $V_{1{\rm R}}$ and $V_{1{\rm L}}$ are given by (\ref{eq:SW}), while for
$m=2$, $V_{2{\rm R}}$ and $V_{2{\rm L}}$ are given by (\ref{eq:V-}). As usual, the
operators $A_m$ and $\Ab_m$ can be written in terms of the superpotentials $W_m$
and are $A_m = \frac{d}{dx} + W_m(x)$, $\Ab_m = - \frac{d}{dx} + W_m(x)$,
$m=1$, 2,~\ldots. For $m=1$, $W_{1{\rm R}}$ and $W_{1{\rm L}}$ are given by
(\ref{eq:W}). In terms of $W_m$, the partner potentials $V^{(\pm)}_m$ read
\begin{equation}
  V^{(\pm)}_m = W_m^2 \mp W_m' + E_{m,0}.
\end{equation}
\par
%
%
We denote the eigenvalues and eigenfunctions of $H_m$ by $E_{m,n}$ and
$\psi_{m,n}$, respectively. These satisfy the SUSY properties
\begin{eqnarray}
  E_{m,n} & = & E_{m-1,n+1} = \cdots = E_{1, m+n-1}, \nonumber \\
  (\psi_{m,n})_{\rm R,L} & = & (C_{m,n})_{\rm R,L} \left[\frac{d}{dx} +
          (W_{m-1})_{R,L}\right] (\psi_{m-1,n+1})_{\rm R,L},
\end{eqnarray}
along with the conditions
\begin{eqnarray}
  (\psi_{m,n})_{\rm R}(0) & = & (\psi_{m,n})_{\rm L}(0) = \alpha_{m,n}, \nonumber \\
  \partial_x (\psi_{m,n})_{\rm R}(0) & = & \partial_x (\psi_{m,n})_{\rm L}(0) =
          {\rm i}\beta_{m,n},  \label{eq:continuity}
\end{eqnarray}
where $\alpha_{m,n}$, $\beta_{m,n} \in \R$. Up to now we have determined the energy
eigenvalues for $m=1$ and 2,
\begin{equation}
  E_{1,n} = E_n = - {\rm i}Z - \kappa_n^2, \qquad E_{2,n} = -{\rm i}Z - \kappa_{n+1}^2,
\end{equation}
with associated eigenfunctions
\begin{eqnarray}
  (\psi_{1,n})_{\rm R} & = & (C_{1,n})_{\rm R} \sinh[\kappa_n (1-x)], \nonumber \\
  (\psi_{1,n})_{\rm L} & = & (C_{1,n})_{\rm L} \sinh[\kappa_n^* (1+x)], \nonumber \\
  (\psi_{2,n})_{\rm R} & = & (C_{2,n})_{\rm R} (C_{1,n+1})_{\rm R} \sinh[\kappa_{n+1}
          (1-x)] \nonumber \\
  && \mbox{} \times \{- \kappa_{n+1} \coth[\kappa_{n+1} (1-x)] + \kappa_0
        \coth[\kappa_0 (1-x)]\}, \nonumber \\
  (\psi_{2,n})_{\rm L} & = & (C_{2,n})_{\rm L} (C_{1,n+1})_{\rm L}
          \sinh[\kappa_{n+1}^* (1+x)] \nonumber \\
  && \mbox{} \times \{\kappa_{n+1}^* \coth[\kappa_{n+1}^* (1+x)] - \kappa_0^*       
          \coth[\kappa_0^* (1+x)]\}.  \label{eq:wf-12}
\end{eqnarray}
In (\ref{eq:wf-12}) we have defined
\begin{eqnarray}
  (C_{1,n})_{\rm R} & = & \frac{\alpha_{1,n}}{\sinh \kappa_n}, \qquad 
          (C_{1,n})_{\rm L} = \frac{\alpha_{1,n}}{\sinh \kappa_n^*}, \nonumber \\
  (C_{2,n})_{\rm R} & = & (C_{2,n})_{\rm L} = C_{2,n} = \frac{\alpha_{2,n}}
          {\alpha_{1,n+1} (- \kappa_{n+1} \coth \kappa_{n+1} + \kappa_0 \coth
          \kappa_0)}.  \label{eq:C} 
\end{eqnarray} 
\par
%
%
To construct the third member of the hierarchy we observe that $W_2(x)$ can be easily
determined from the property $A_2 \psi_{2,0} = 0$ that yields $W_2(x) = - \psi'_{2,0}/
\psi_{2,0}$. Explicit calculations give for $W_2$  the forms
\begin{eqnarray}
  W_{2,{\rm R}}(x) & = & - \kappa_0 \coth[\kappa_0 (1-x)] + \frac{\kappa_1^2 -
         \kappa_0^2}{\kappa_1 \coth[\kappa_1 (1-x)] - \kappa_0 \coth[\kappa_0
         (1-x)]}, \nonumber \\
  W_{2,{\rm L}}(x) & = & - W_{2{\rm R}}^*(-x). \label{eq:W-2}
\end{eqnarray}
\par
%
%
{}For the potential $V_3(x)$, our results turn out to be
\begin{eqnarray}
  V_{3,{\rm R}}(x) & = & - {\rm i}Z - 2(\kappa_1^2 - \kappa_0^2) \frac{\kappa_1^2
          \cosech^2[\kappa_1 (1-x)] - \kappa_0^2 \cosech^2[\kappa_0 (1-x)]}
          {\{\kappa_1 \coth[\kappa_1 (1-x)] - \kappa_0 \coth[\kappa_0 (1-x)]\}^2},
          \nonumber \\ 
  V_{3,{\rm L}}(x) & = & V_{3,{\rm R}}^*(-x), \label{eq:V-3} 
\end{eqnarray}
while for the eigenfunctions we get
\begin{eqnarray}
  (\psi_{3,n})_{\rm R} & = & (C_{3,n})_{\rm R} (C_{2,n+1})_{\rm R} (C_{1,n+2})_{\rm
        R} \sinh[\kappa_{n+2} (1-x)]  \Biggl\{\kappa_{n+2}^2 - \kappa_0^2  \nonumber \\
  && \mbox{} - (\kappa_1^2 - \kappa_0^2) \frac{\kappa_{n+2} \coth[\kappa_{n+2}
        (1-x)] - \kappa_0 \coth[\kappa_0 (1-x)]}{\kappa_1 \coth[\kappa_1 (1-x)] -
        \kappa_0 \coth[\kappa_0 (1-x)]}\Biggr\}, \nonumber \\ 
  (\psi_{3,n})_{\rm L} & = & (C_{3,n})_{\rm L} (C_{2,n+1})_{\rm L} (C_{1,n+2})_{\rm
        L} \sinh[\kappa_{n+2}^* (1+x)]  \Biggl\{\kappa_{n+2}^{*2} - \kappa_0^{*2} 
        \nonumber  \\
  && \mbox{} - (\kappa_1^{*2} - \kappa_0^{*2}) \frac{\kappa_{n+2}^*
        \coth[\kappa_{n+2}^* (1+x)] - \kappa_0^* \coth[\kappa_0^* (1+x)]}{\kappa_1^*
        \coth[\kappa_1^* (1+x)] - \kappa_0^* \coth[\kappa_0^* (1+x)]}\Biggr\},  
        \label{eq:wf-3}
\end{eqnarray}
where $E_{3,n} = - {\rm i}Z - \kappa_{n+2}^2$ and $(\psi_{3,n})_{\rm R}(1) =
(\psi_{3,n})_{\rm L}(-1) = 0$.\par
%
%
It remains now to impose the continuity conditions (\ref{eq:continuity}) on the
eigenfunctions (\ref{eq:wf-3}). While the first one leads to
\begin{eqnarray}
  \lefteqn{(C_{3,n})_{\rm R} = (C_{3,n})_{\rm L} = C_{3,n} \nonumber }\\
  & = & \frac{\alpha_{3,n}}{\alpha_{2,n+1}}  \biggl\{\frac{\kappa_1^2 -
        \kappa_0^2}{\kappa_1 \coth\kappa_1 - \kappa_0 \coth \kappa_0} -
        \frac{\kappa_{n+2}^2 - \kappa_0^2}{\kappa_{n+2} \coth\kappa_{n+2}
        - \kappa_0 \coth \kappa_0}\biggr\}^{-1},
\end{eqnarray}
the second one amounts to an identity when (\ref{eq:C}) and (\ref{eq:wf-3}) are taken
into account.\par
%
%
We have thus obtained explicit forms of the first three members of the SUSY hierarchy for
the PT-symmetric square well potential. It is clear that by applying similar techniques,
formulas for other members may be similarly constructed.\par
%
%
It is easy to check that in the $Z \to 0$ limit, all the results obtained in this section go
over to those for the third member of the real square well hierarchy. Equations
(\ref{eq:W-2}) and (\ref{eq:V-3}), for instance, yield $W_{2,{\rm R}}(x) = W_{2,{\rm
L}}(x) = W_2(x) = \pi \tan\left(\frac{\pi}{2} x\right)$ and $V_{3,{\rm R}}(x) =
V_{3,{\rm L}}(x) = V_3(x) = \frac{3}{2} \pi^2 \sec^2\left(\frac{\pi}{2} x\right)$, in
conformity with Eq.~(\ref{eq:V-m}).\par
%
%
It is useful to stress here that, in the same limit, the eigenfunctions of the first three
members in the hierarchy, namely (\ref{eq:wf}), (\ref{eq:partner-wf}), and
(\ref{eq:wf-3}), turn out to be proportional to Gegenbauer polynomials:
\begin{eqnarray}
  Z \to 0: \qquad \psi_{1,n}(x) & = & - {\rm i} C_{1,n} \cos\left(\frac{\pi}{2} x\right)
       C^{(1)}_n\left[\sin \left(\frac{\pi}{2} x\right)\right], \nonumber \\
  \psi_{2,n}(x) & = & - {\rm i} \pi C_{2,n} C_{1,n+1} \cos^2\left(\frac{\pi}{2} x\right)
       C^{(2)}_n\left[\sin \left(\frac{\pi}{2} x\right)\right], \nonumber \\ 
  \psi_{3,n}(x) & = & - 2{\rm i} \pi^2 C_{3,n} C_{2,n+1} C_{1,n+2}
       \cos^3\left(\frac{\pi}{2} x\right) C^{(3)}_n\left[\sin \left(\frac{\pi}{2}
x\right)\right]. 
       \label{eq:gegenbauer}
\end{eqnarray}
In (\ref{eq:wf-real}) and (\ref{eq:partner-wf-real}) we had already furnished the limiting
forms ($Z \to 0$) of the first two members in the hierarchy. Note that to get to the
representations (\ref{eq:gegenbauer}) we used the definition of the Gegenbauer
polynomial $C^{(1)}_n(\cos\phi) = [\sin(n+1)\phi]/\sin\phi$ and considered the general
recurrence relation
\begin{eqnarray}
  &&\left[\frac{d}{dx} + (m-1) \frac{\pi}{2} \tan\left(\frac{\pi}{2} x\right)\right]
        \cos^{m-1}\left(\frac{\pi}{2} x\right) C^{(m-1)}_{n+1}\left[\sin \left(\frac{\pi}{2}
        x\right)\right] \\
  && = \pi (m-1) \cos^m\left(\frac{\pi}{2} x\right) C^{(m)}_n
        \left[\sin \left(\frac{\pi}{2} x\right)\right], 
\end{eqnarray}
which can be easily obtained from known properties of Gegenbauer
polynomials~\cite{erdelyi}. For nonvanishing values of Z, it can be shown that the
eigenfunctions of the PT-symmetric square well and of the next two members in the
hierarchy, namely (\ref{eq:wf-12}) and (\ref{eq:wf-3}), can be rewritten in terms of
Gegenbauer functions of the type $C^{(p)}_{(\kappa_{n+q}/\kappa_0)-p}
\{\cosh[\kappa_0 (1-x)]\}$ or $C^{(p)}_{(\kappa_{n+q}^*/\kappa_0^*)-p}
\{\cosh[\kappa_0^* (1+x)]\}$, where $p, q \in \N$.\par
%
%
\section{SUSY Hierarchies in the Broken PT-Symmetry Case}

Let us now consider coupling strengths $Z$ for which PT symmetry is spontaneously
broken and assume first that $Z$ lies between the first two critical values, $Z_0^{(\rm
crit)} \approx 4.48 < Z < Z_1^{(\rm crit)} \approx 12.80$~\cite{znojil01b}. The
PT-symmetric square well has then a single pair of complex-conjugate eigenvalues $E_0 =
e_0 - {\rm i}\epsilon_0$, $E_1 = E_0^* = e_0 + {\rm i}\epsilon_0$ (where $\epsilon_0
> 0$), and an infinite sequence of real eigenvalues $E_n$, $n=2$, 3,~\ldots.\par
%
%
The corresponding Schr\"odinger equation is equivalent to
\begin{equation}
  \psi''_{n{\rm R}} = \rho_n^2 \psi_{n{\rm R}}, \qquad \psi''_{n{\rm L}} = \sigma_n^2
  \psi_{n{\rm L}},  \label{eq:SE-broken}  
\end{equation}
where
\begin{eqnarray}
  \rho_0 & = & \kappa_0, \qquad \rho_1 = \lambda_0, \qquad \rho_n = \kappa_n,
        \qquad n=2, 3, \ldots, \nonumber\\
  \sigma_0 & = & \lambda_0^*, \qquad \sigma_1 = \kappa_0^*, \qquad \sigma_n =
        \kappa_n^*, \qquad n=2, 3, \ldots,
\end{eqnarray}
and
\begin{equation}
  \kappa_0^2 = - e_0 + {\rm i}\epsilon_0 - {\rm i}Z, \qquad \lambda_0^2 = - e_0 - {\rm
  i}\epsilon_0 - {\rm i}Z, \qquad \kappa_n^2 = - E_n - {\rm i}Z, \qquad n=2, 3, \ldots.
\end{equation}
\par
%
%
The solutions of (\ref{eq:SE-broken}), vanishing at the end points of the interval $(-1,
+1)$ and satisfying the continuity conditions of $\psi_n$ and $\psi'_n$ at $x=0$, are
given by
\begin{equation}
  \psi_{n{\rm R}} = C_{n{\rm R}} \sinh[\rho_n(1-x)], \qquad \psi_{n{\rm L}} = C_{n{\rm
  L}} \sinh[\sigma_n(1+x)], 
\end{equation}
where
\begin{equation}
  \rho_n \coth\rho_n + \sigma_n \coth\sigma_n = 0, \qquad \frac{C_{n{\rm R}}}
  {C_{n{\rm L}}} = \frac{\sinh\sigma_n}{\sinh\rho_n}.
\end{equation}
Choosing them real at $x=0$, we obtain
\begin{equation}
  C_{n{\rm R}} = \frac{\alpha_n}{\sinh\rho_n}, \qquad C_{n{\rm L}} =
  \frac{\alpha_n}{\sinh\sigma_n}, \qquad \alpha_n \in \R. 
\end{equation}
With this choice, the eigenfunctions $\psi_n$, $n=2$, 3,~\ldots, corresponding to the
real eigenvalues $E_n$, $n=2$, 3,~\ldots, are PT-symmetric as in the unbroken
PT-symmetry case, but this does not hold true for $\psi_0$, $\psi_1$, corresponding to
the complex-conjugate eigenvalues $E_0$, $E_1$.\par
%
%
To construct a SUSY hierarchy of partner Hamiltonians $H_m$, $m=1$, 2,~\ldots, whose
first member coincides with that of the PT-symmetric square well, we have now various
possibilities at our disposal. At each stage, we may indeed eliminate either the lowest real
eigenvalue or one of the two complex eigenvalues as long as there remains some. As we
shall proceed to show, this leads to a rich diversity of hierarchies: a PT-symmetric
hierarchy with spontaneous symmetry breaking, two non-PT-symmetric hierarchies with
partial PT-symmetry restoration, as well as various mixed-type hierarchies.\par
%
%
Let us start with the first one to be distinguished by an upper index (1). By successively
eliminating the lowest-lying real eigenvalues $E^{(1)}_{1,2} = E_2$, $E^{(1)}_{1,3} =
E_3$,~\ldots, while keeping the two complex ones $E^{(1)}_{1,0} = E_0$,
$E^{(1)}_{1,1} = E_1$ at every stage, we obtain PT-symmetric potentials
$V^{(1)}_m(x)$, $m=1$, 2, 3,~\ldots, with a pair of complex-conjugate eigenvalues 
$E^{(1)}_{m,0} = E_0$, $E^{(1)}_{m,1} = E_1$, and an infinite sequence of real
eigenvalues $E^{(1)}_{m,2} = E_{m+1}$, $E^{(1)}_{m,3} = E_{m+2}$,~\ldots. The
corresponding eigenfunctions $\psi^{(1)}_{m,n}(x)$, $n=0$, 1, 2,~\ldots, turn out to
be PT-symmetric only starting from $n=2$.\par
%
%
This is illustrated with the second and third members of the hierarchy, which can be
obtained from the results of Sec.~5 by making some appropriate substitutions:
$\kappa_0 \to \rho_2$, $\kappa_1 \to \rho_3$, $\kappa_0^* \to \sigma_2$,
$\kappa_1^* \to \sigma_3$ for $W^{(1)}_1$, $V^{(1)}_2$, $W^{(1)}_2$, and
$V^{(1)}_3$, and the same together with $\kappa_1 \to \rho_0$, $\kappa_2 \to
\rho_1$, $\kappa_1^* \to \sigma_0$, $\kappa_2^* \to \sigma_1$ or $\kappa_2 \to
\rho_0$, $\kappa_3 \to \rho_1$, $\kappa_2^* \to \sigma_0$,
$\kappa_3^* \to \sigma_1$ for  $\psi^{(1)}_{2,n}$ and $\psi^{(1)}_{3,n}$,
respectively. The corresponding potentials, for instance, are given by
\begin{eqnarray}
  V^{(1)}_{2,{\rm R}}(x) & = & - {\rm i}Z + 2 \kappa_2^2 \cosech^2[\kappa_2 (1-x)],
        \nonumber \\
  V^{(2)}_{2,{\rm L}}(x) & = & V^{(1)*}_{2,{\rm R}}(-x),
\end{eqnarray}
and
\begin{eqnarray}
  V^{(1)}_{3,{\rm R}}(x) & = & - {\rm i}Z - 2(\kappa_3^2 - \kappa_2^2)
          \frac{\kappa_3^2 \cosech^2[\kappa_3 (1-x)] - \kappa_2^2 \cosech^2[\kappa_2
          (1-x)]}{\{\kappa_3 \coth[\kappa_3 (1-x)] - \kappa_2 \coth[\kappa_2
          (1-x)]\}^2}, \nonumber \\ 
  V^{(1)}_{3,{\rm L}}(x) & = & V^{(1)*}_{3,{\rm R}}(-x). 
\end{eqnarray}
We conclude that all the numbers of this first hierarchy have properties very similar to
those of the PT-symmetric square well for the chosen $Z$ value, namely all of them are
PT-symmetric but exhibit spontaneously broken PT symmetry with a single pair of
complex-conjugate eigenvalues.\par
%
%
Let us now consider the second type of hierarchies, which are obtained by successively
eliminating the two complex eigenvalues, then the real ones in increasing energy order.
Since we may choose to eliminate first $E_0$, then $E_1$, or the reverse, there are two
different hierarchies to be referred to by an upper index (2) or (3), respectively.\par
%
%
To obtain results for the former from those of Sec.~5, it is enough to perform the
substitutions $\kappa_n \to \rho_n$, $\kappa_n^* \to \sigma_n$. So we get for
instance
\begin{eqnarray}
  V^{(2)}_{2,{\rm R}}(x) & = & - {\rm i}Z + 2 \kappa_0^2 \cosech^2[\kappa_0 (1-x)],
        \nonumber \\
  V^{(2)}_{2,{\rm L}}(x) & = & {\rm i}Z + 2 \lambda_0^{*2} \cosech^2[\lambda_0^*
        (1+x)],
\end{eqnarray}
and
\begin{eqnarray}
  V^{(2)}_{3,{\rm R}}(x) & = & - {\rm i}Z - 2(\lambda_0^2 - \kappa_0^2)
          \frac{\lambda_0^2 \cosech^2[\lambda_0 (1-x)] - \kappa_0^2
          \cosech^2[\kappa_0 (1-x)]}{\{\lambda_0 \coth[\lambda_0 (1-x)] - \kappa_0
          \coth[\kappa_0 (1-x)]\}^2}, \nonumber \\ 
  V^{(2)}_{3,{\rm L}}(x) & = & V^{(2)*}_{3,{\rm R}}(-x), 
\end{eqnarray}
while the remaining potentials $V^{(2)}_n(x)$, $n=4$, 5,~\ldots, are PT-symmetric. The
potential $V^{(2)}_2(x)$ and the eigenfunctions of the corresponding Hamiltonians
$H^{(2)}_2$ are not PT-symmetric, which is not surprising~\cite{bender, kretschmer,
bagchi01b, japaridze, mosta, ahmed} since $H^{(2)}_2$ has a single  complex eigenvalue
$E^{(2)}_{2,0} = E_1$ in addition to real ones, $E^{(2)}_{2,n} = E_{n+1}$, $n=1$,
2,~\ldots. Strangely enough, the next potential $V^{(2)}_3$ is PT-symmetric and the
corresponding spectrum $E^{(2)}_{3,n} = E_{n+2}$, $n=0$, 1, 2,~\ldots, is entirely
real, {\em but} the eigenfunctions $\psi^{(2)}_{3,n}(x)$, with no definite symmetry
under PT, do not even differ by a simple phase factor from $\psi^{(2)*}_{3,n}(-x)$, as it
is normally the case for eigenfunctions of PT-symmetric Hamiltonians~\cite{bagchi01b}.
This distinctive feature may be traced back to the finite discontinuity of the
PT-symmetric square well and of its SUSY partners at $x=0$, which has led us to ensure
smoothness of the eigenfunctions by imposing continuity conditions externally by hand.
As explicitly shown here, such a procedure does not guarantee PT symmetry of the
eigenfunctions at all levels. In contrast, the results of~\cite{bagchi01b} were based on
the tacit assumption of smoothness of the potential over the entire real line and may
therefore not be compared with the present situation.\par
%
%
The other second-type hierarchy only differs from the first one by the interchange of
$E_0$ and $E_1$ or $\kappa_0$, $\kappa_0^*$ and $\lambda_0$, $\lambda_0^*$. As
a result, we get
\begin{equation}
  V^{(3)}_2(x) = V^{(2)*}_2(-x), \qquad V^{(3)}_3(x) = V^{(2)}_3(x).
\end{equation}
Hence all the corresponding potentials of the two hierarchies coincide, but for the second
ones, which are related through PT symmetry. It can also be shown that from $m=2$
onwards, the eigenfunctions of corresponding members $H^{(2)}_m$, $H^{(3)}_m$ of
the two hierarchies are also related through PT symmetry: $\psi^{(3)}_{m,n}(x) =
\psi^{(2)*}_{m,n}(-x)$, $m=2$, 3,~\ldots, $n=0$, 1, 2,~\ldots. We therefore conclude
that these second-type hierarchies are non-PT-symmetric ones, but exhibit partial
PT-symmetry restoration.\par
%
%
It should now be clear that apart from the hierarchies considered so far, there also exist a
lot of mixed-type ones, which differ from one another and from the previous ones in the
step (resp.\ steps) where one of the two (resp.\ both) complex eigenvalues is (resp.\
are) eliminated.\par
%
%
{}Finally, if the coupling strength $Z$ lies in another interval $\left(Z_{\nu-1}^{\rm (crit)},
Z_{\nu}^{\rm (crit)}\right)$, such that $\nu \in {2, 3, \ldots}$, and there therefore exist
$\nu$ pairs of complex-conjugate eigenvalues ($E_0$, $E_1 = E_0^*$), ($E_2$, $E_3 =
E_2^*$), \ldots, ($E_{2\nu-2}$, $E_{2\nu-1} = E_{2\nu-2}^*$) in addition to the real
ones, SUSY hierarchies of partner Hamiltonians can be constructed in many different
ways. In particular, by eliminating the complex eigenvalues, it is possible to partially restore
PT symmetry after $2\nu$ steps.\par
%
%
\section{Conclusion}

In the present Letter, we have established the SUSY connection of the PT-symmetric
square well both in the unbroken and broken PT-symmetry cases. In the former we have
availed ourselves of this to derive a PT-symmetric analogue of the $\sec^2$-hierarchy.\par
%
%
In this respect, the PT-symmetric world proves more intricate than the Hermitian one: not
only has one to resort to numerical calculations to determine the eigenvalues, but also
the potentials and the eigenfunctions get more and more complicated when going to
successive members of the hierarchy in contrast to what happens for the real square
well.\par
%
%
Another intricacy of the PT-symmetric square well problem, namely the existence of
complex eigenvalues for a coupling strength above the critical threshold
$Z^{\rm(crit)}_0$, leads to a new feature in the SUSY extension to non-Hermitian
Hamiltonians: the existence of a rich diversity of SUSY hierarchies, including some with
PT-symmetry breaking and some with partial PT-symmetry restoration.\par
%
%
\section*{Acknowledgments}

One of us (S.\ M.) thanks the Council of Scientific and Industrial Research, New Delhi for
financial support.\par
%
%


\newpage
\begin{thebibliography}{99}

\bibitem{schiff} L.\ I.\ Schiff, {\em Quantum Mechanics} (McGraw-Hill, New York, 1968).

\bibitem{sukumar} C.\ V.\ Sukumar, {\em J.\ Phys.} {\bf A18}, L57 (1985).

\bibitem{bender} C.\ M.\ Bender and S.\ Boettcher, {\em Phys.\ Rev.\ Lett.} {\bf 80},
5243 (1998).

\bibitem{fernandez} F.\ M.\ Fern\'andez, R.\ Guardiola, J.\ Ros and M.\ Znojil, {\em J.\
Phys.} {\bf A31}, 10105 (1998).

\bibitem{cannata} F.\ Cannata, G.\ Junker and J.\ Trost, {\em Phys.\ Lett.} {\bf A246},
219 (1998).

\bibitem{delabaere} E.\ Delabaere and F.\ Pham, {\em Phys.\ Lett.} {\bf A250}, 25, 29
(1998).

\bibitem{bagchi00a} B.\ Bagchi, F.\ Cannata and C.\ Quesne, {\em Phys.\ Lett.} {\bf
A269}, 79 (2000).

\bibitem{bagchi00b} B.\ Bagchi and C.\ Quesne, {\em Phys.\ Lett.} {\bf A273}, 285
(2000).

\bibitem{andrianov} A.\ A.\ Andrianov, M.\ V.\ Ioffe, F.\ Cannata and J.-P.\ Dedonder,
{\em Int.\ J.\ Mod.\ Phys.} {\bf A14}, 2675 (1999).

\bibitem{bagchi00c} B.\ Bagchi and R.\ Roychoudhury, {\em J.\ Phys.} {\bf A33}, L1
(2000).

\bibitem{znojil00} M.\ Znojil, F.\ Cannata, B.\ Bagchi and R.\ Roychoudhury, {\em Phys.\
Lett.} {\bf B483}, 284 (2000).

\bibitem{bagchi01a} B.\ Bagchi, S.\ Mallik and C.\ Quesne, {\em Int.\ J.\ Mod.\ Phys.}
{\bf A16}, 2859 (2001).

\bibitem{bessis} D.\ Bessis, unpublished (1992).

\bibitem{kretschmer} R.\ Kretschmer and L.\ Szymanowski, ``The interpretation of
quantum-mechanical models with non-Hermitian Hamiltonians and real spectra'',
quant-ph/0105054.

\bibitem{bagchi01b} B.\ Bagchi, C.\ Quesne and M.\ Znojil, {\em Mod.\ Phys.\ Lett.}
{\bf A16}, 2047 (2001).

\bibitem{ahmed} Z.\ Ahmed, {\em Phys.\ Lett.} {\bf A282}, 343 (2001); {\bf A287},
295 (2001).

\bibitem{mosta} A.\ Mostafazadeh, {\em J.\ Math.\ Phys.} {\bf 43}, 205 (2002).

\bibitem{japaridze} G.\ S.\ Japaridze, {\em J.\ Phys.} {\bf A35}, 1709 (2002).

\bibitem{znojil01a} M.\ Znojil, {\em Phys.\ Lett.} {\bf A285}, 7 (2001).

\bibitem{znojil01b} M.\ Znojil and G.\ L\'evai, {\em Mod.\ Phys.\ Lett.} {\bf A16}, 2273
(2001).

\bibitem{erdelyi} A.\ Erd\'elyi, W.\ Magnus, F.\ Oberhettinger and F.\ G.\ Tricomi, {\em
Higher Transcendental Functions} (McGraw-Hill, New York, 1953), Vol.\ I.

\end {thebibliography}
\end{document}